\documentclass{mn2e}
\usepackage{graphicx}
\newcommand{\bT}{\bmath{T}}

\newcommand{\text}[1]{\quad\mbox{#1}\quad}

\newcommand{\sub}[1]{_{\mbox{\tiny #1}}}

\title{Stellar explosions powered by the Blandford-Znajek mechanism}

\author[M.V.Barkov \& S.S. Komissarov ]{
Maxim V.~Barkov,$^{1,2}$\footnotemark[1]
Serguei S.~Komissarov,$^{1}$\thanks{ 
email:~bmv@maths.leeds.ac.uk;
serguei@maths.leeds.ac.uk}\\
$^{1}$Department of Applied Mathematics, The University of Leeds,
Leeds, LS2 9GT\\
$^{2}$Space Research Institute, 84/32 Profsoyuznaya Street, Moscow
117997, Russia}
\begin{document}
\date{Received/Accepted}
\maketitle

\begin{abstract}
In this letter we briefly describe the first results of our numerical study
on the possibility of magnetic origin of relativistic jets of long duration 
gamma ray bursters within the collapsar scenario. 
We track the collapse of massive rotating stars onto a rotating central black 
hole using axisymmetric general relativistic magnetohydrodynamic code that 
utilizes a realistic equation of state of stellar matter,
takes into account the cooling associated with emission of neutrinos,
and the energy losses due to dissociation of nuclei. The neutrino heating
is not included.  
We describe the solution for one particular model where the progenitor star 
has magnetic field $B=3\times10^{10}$G. The solution exhibits strong explosion 
driven by the Poynting-dominated jets whose power exceeds 
$2\times10^{51}\,\mbox{erg/s}$. The jets originate mainly from the black 
hole and they are powered via the Blandford-Znajek mechanism. 
\end{abstract}
                                                                                          
                                                                                          
\section{Introduction}
\label{introduction}
The phenomenon of Gamma Ray Burst (GRB) has been puzzling
astrophysicists for many years since its discovery in
1970s~\cite{KSO73,MGI74}. The recent identification of long duration
GRBs with supernovae (see Della Valle 2006, and Woosley \& Bloom 2006
for full review) means that we are dealing with enormous amount of
energy, $10^{51}-10^{52}\mbox{erg}$, released within a very short
time, 2-100 seconds, in the form of highly relativistic collimated
outflow \cite{P05}.  Most of the current GRB studies are focused on
the physics associated with production of gamma rays in such flows and
their interaction with the interstellar medium or the stellar wind of
the supernova progenitor. However, the central question in the problem
of GRBs is undoubtedly the nature of their central engines. These
powerful jets have to be produced as a result of stellar collapse,
most likely by the relativistic object, neutron star or black hole
(BH), formed in the center, and make their way through the massive
star unscathed, remaining well collimated and highly relativistic.

The most popular model of central engine is based on the ``failed
supernova'' scenario of stellar collapse, or ``collapsar'', where the
iron core of progenitor star forms a BH \cite{W93}.  If the progenitor
is non-rotating then its collapse is likely to continue in a
``silent'' manner until the whole star is swallowed by the BH.  If,
however, the specific angular momentum in the equatorial part of
stellar envelope exceeds that of the last stable orbit of the BH then
the collapse becomes highly anisotropic.  While in the polar region it
may proceed more or less uninhibited, for a while, the equatorial
layers form dense and massive accretion disk. The gravitational energy
released in the disk can be very large, more then sufficient to stop
the collapse of outer layers and drive GRB outflows, presumably in the
polar direction where density is much lower \cite{MW99}. In addition,
there is plenty of rotational energy in the BH itself

\begin{equation} 
  E\sub{rot} = \frac{M\sub{bh}c^2}{2} \left\{ 2-\left[
  \left(1+\sqrt{1-a^2}\right)^2+a^2 \right]^{1/2} \right\},
\end{equation}   
where $M\sub{bh}$ is the BH mass and $a\in(-1,1)$ is its dimensionless
rotation parameter. For $M\sub{bh}=3M_{\sun}$ and $a=0.9$ this gives
the enormous value of $E\sub{rot} \simeq8\times10^{53}$erg.

The three currently actively discussed mechanisms of powering GRB jets
in the collapsar scenario are the heating via annihilation of
neutrinos produced in the disk \cite{MW99}, the magnetic braking of
the disk \cite{BP82,UM06}, and the magnetic braking of the BH
\cite{BZ77}.  The potential role of neutrino mechanism is rather
difficult to assess as this requires accurate treatment of neutrino
transport in a complex dynamic environment of collapsar. The long and
complicated history of numerical studies of neutrino-driven supernova
explosions teaches us to be cautious.  Numerical simulations by
MacFadyen \& Woosley\shortcite{MW99} and Aloy et
al.\shortcite{AIMGM00} have demonstrated that sufficiently large
energy deposition in the polar region above the disk may indeed result
in fast collimated jets. However, the neutrino transport has not been
implemented in these simulations and the energy deposition was based
simply on expectations.  When Nagataki et al.\shortcite{NTMT07}
utilized a simple prescription for neutrino transport in their code
they found that neutrino heating was insufficient to drive polar jets.

A number of groups have studied the collapsar scenario using Newtonian
MHD codes and implementing the Paczynski-Witta potential in order to
approximate the gravitational field of central BH
\cite{PMAB03,FKYHS06,NTMT07}. In this approach it is impossible to
capture the Blandford-Znajek effect and only the magnetic braking of
the accretion disk can be investigated. The general conclusion of
these studies is that the accretion disk can launch
magnetically-driven jets provided the magnetic field in the progenitor
core is sufficiently strong.
Unfortunately, the jet power has not been given in most of these
papers and is difficult to evaluate from the published numbers.  In
the simulations of Proga et al.\shortcite{PMAB03} the jet power at
$t\simeq 0.25$s is $\simeq 10^{50}\mbox{erg}/$s.  The initial magnetic
field in these simulations is monopole with $B\simeq 2\times 10^{14}$G
at $r=3r_g$, where $r_g=GM\sub{bh}/c^2$ (private communication).

\begin{figure*}
\includegraphics[width=57mm]{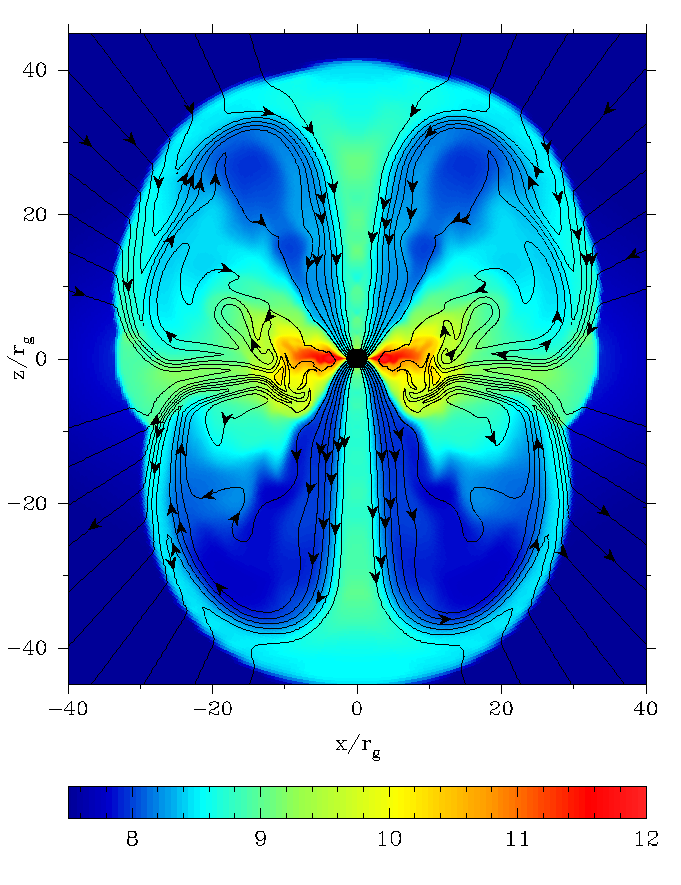}
\includegraphics[width=57mm]{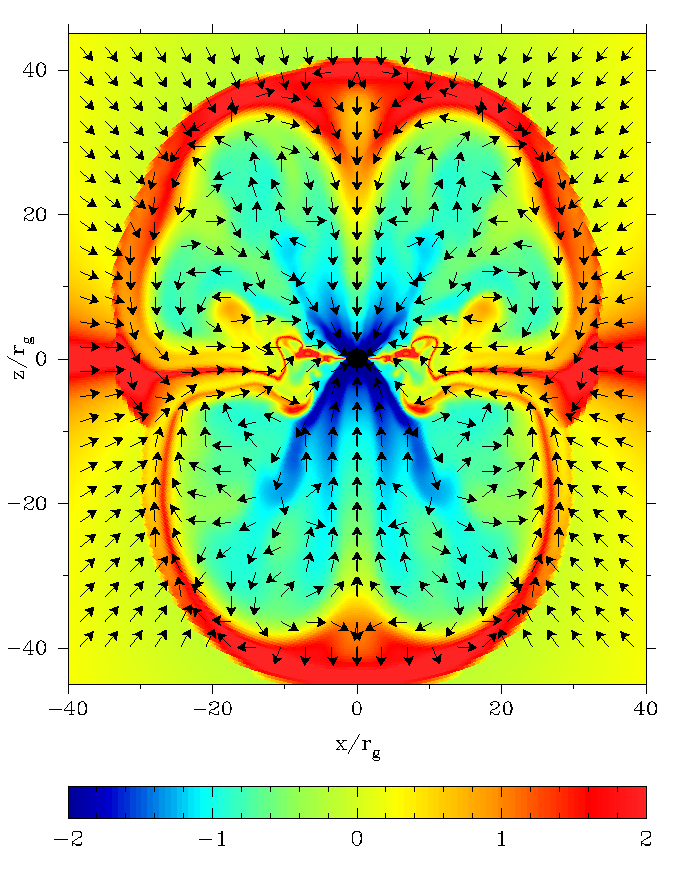}
\includegraphics[width=57mm]{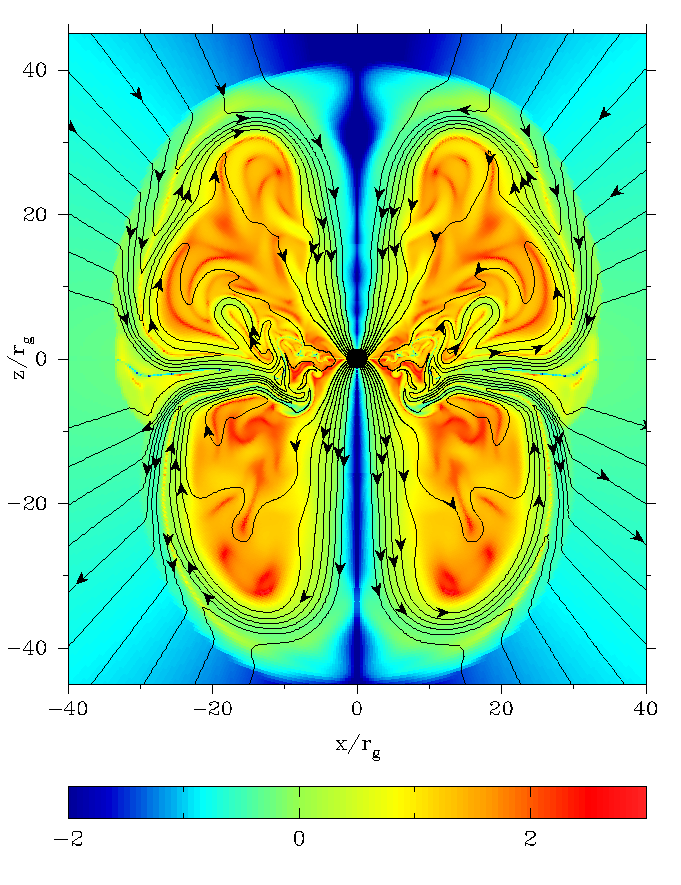}
\caption{Solution immediately before the explosion (t=0.24s).  Left
panel: the baryonic rest mass density, $log_{10}\rho$, in $g/cm^3$ and
the magnetic field lines; Middle panel: the ratio of gas and magnetic
pressures, $log_{10}P/P_m$, and velocity direction vectors; Right
panel: the ratio of azimuthal and poloidal magnetic field strengths,
$log_{10}B^\phi/B_p$, and the magnetic field lines.  }
\label{f0}
\end{figure*}

The study of collapsars in full GRMHD is still in its infancy.
Sekiguchi \& Shibata\shortcite{SS07} studied the collapse of rotating
stellar cores and formation of BH in the collapsar scenario. Their
results show powerful explosions soon after the accretion disk is
formed around the BH and the free falling plasma of polar regions
collides with this disk. These explosions are driven by the heat
generated as a result of such collision. However, the authors have not
accounted for the neutrino cooling and the energy losses due to
photo-dissociation of atomic nuclei.
and the explosions could be similar in nature to the ``successful''
prompt explosions of early supernova simulations \cite{bethe}. Mizuno
et al.\shortcite{MYKS04a,MYKS04b} carried out GRMHD simulations in the
time-independent space-time of a central BH.  The computational domain
did not include the BH ergosphere and thus they could not study the
role of the Blandford-Znajek effect~\cite{K04a}. The energy losses
have not been included and the equation of state (EOS) was a simple
polytrope. These simulations were run for a rather short time, $\simeq
280 r_g/c$ where $r_g=GM/c^2$, and jets were formed almost immediately
due to unrealistically strong initial magnetic field.

In this letter we describe the first results of axisymmetric GRMHD
simulations of collapsars where we use realistic EOS~\cite{TS00},
include the energy losses due to neutrino emission (assuming optically
thin regime) and photo-dissociation of nuclei (see the details of
micro-physics in Komissarov \& Barkov 2007 ), use the computational
domain that includes the BH horizon and its ergosphere, and run
simulations for a relatively long physical time, up to 0.5s. The
neutrino heating is not included.

\begin{figure*}
\includegraphics[width=57mm]{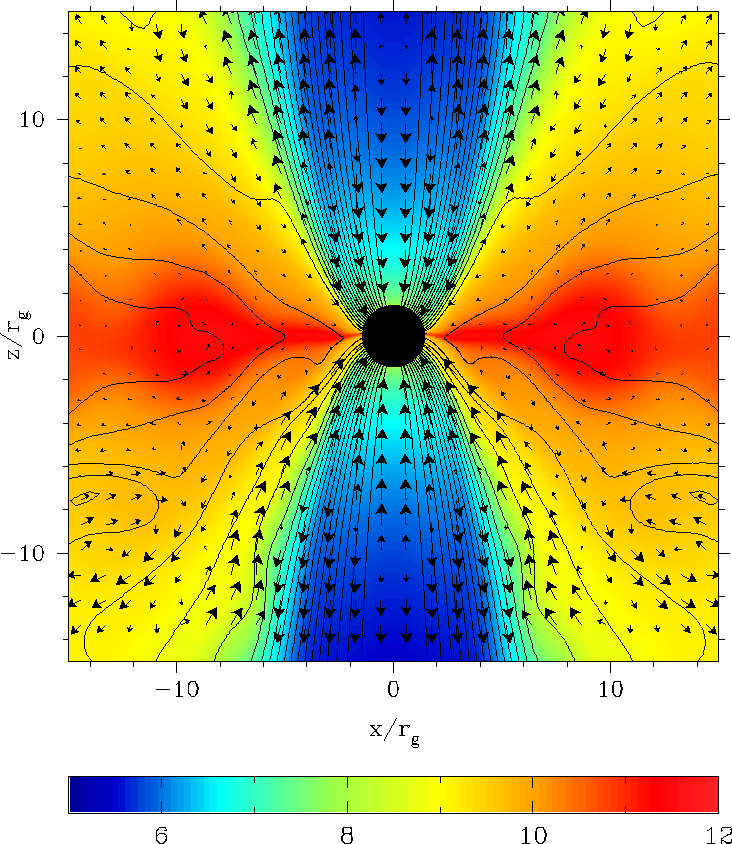}
\includegraphics[width=57mm]{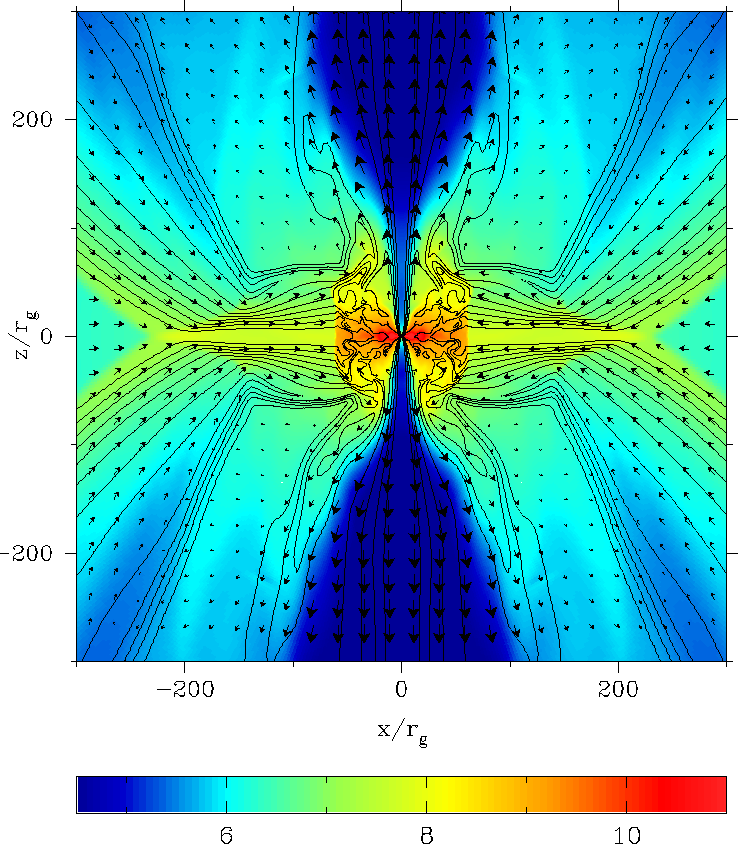}
\includegraphics[width=57mm]{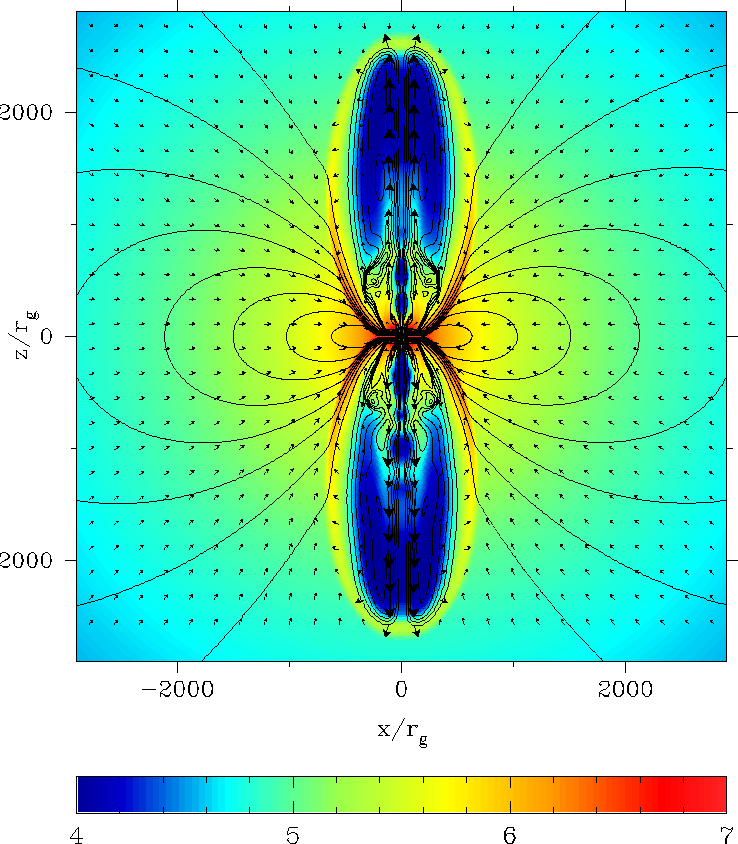}
\caption{Solution on different scales at $t=0.45$s. The colour images
show the baryonic rest mass density, $log_{10}\rho$ in g/cm$^3$, the
contours show the magnetic field lines, and the arrows show the
velocity field.}
\label{f1}
\end{figure*}

\begin{figure*}
\includegraphics[width=57mm]{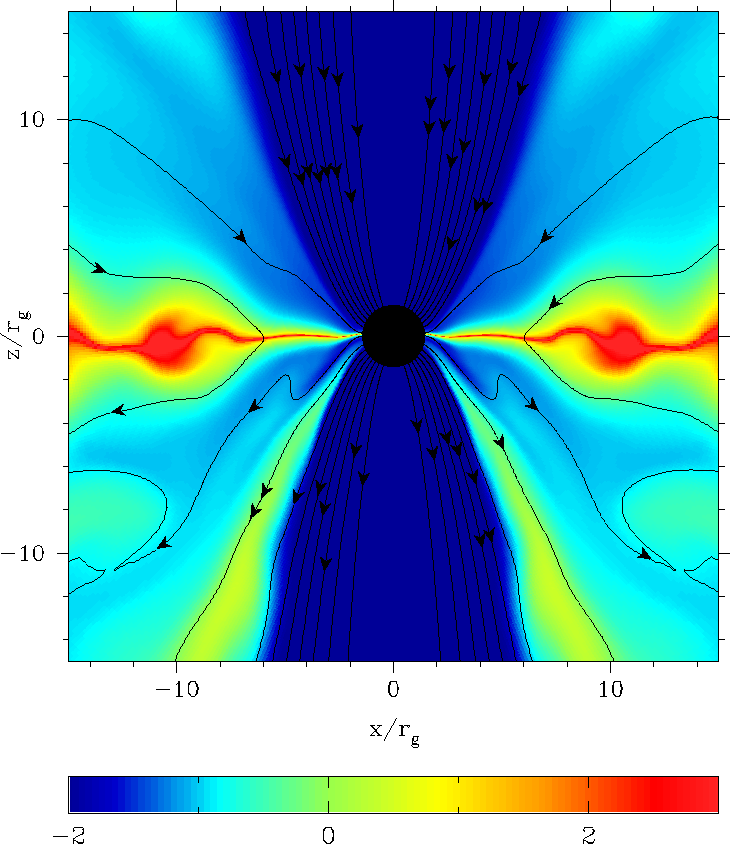}
\includegraphics[width=57mm]{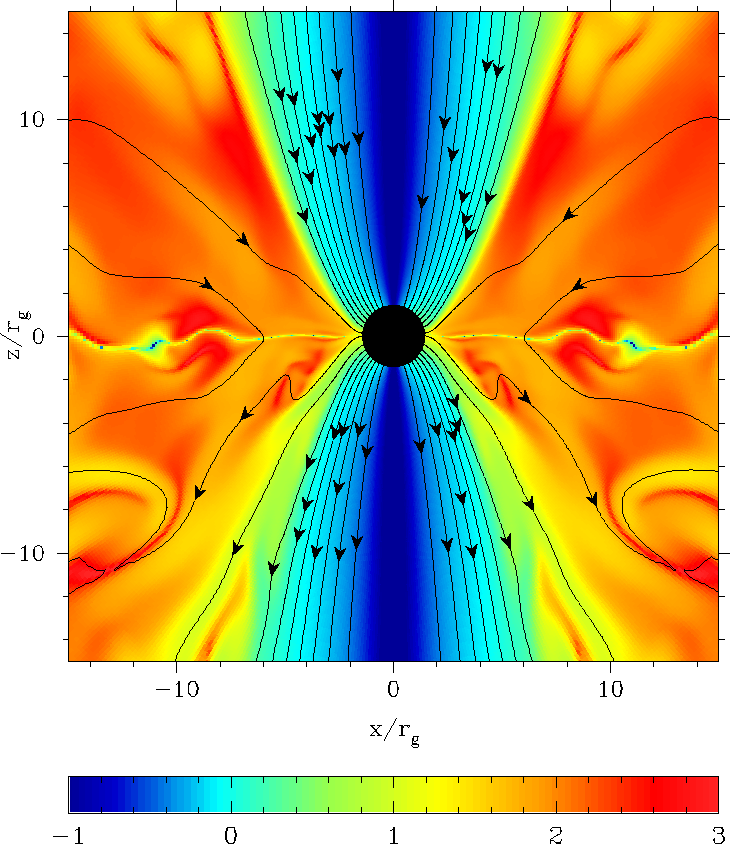}
\includegraphics[width=57mm]{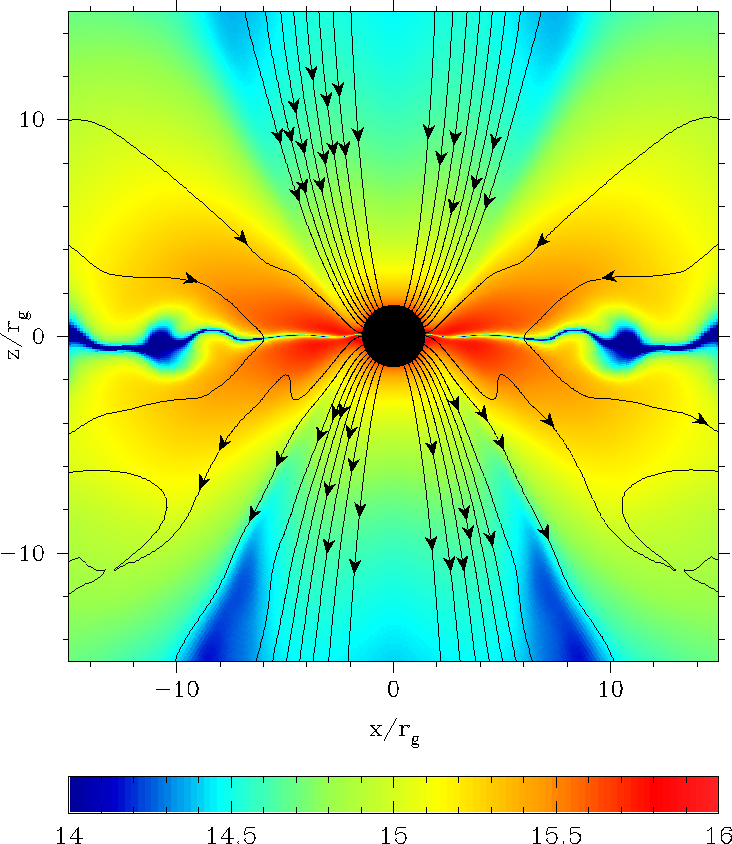}
\caption{The inner region at $t=0.45$s.  Left panel: the magnetization
parameter, $log_{10}P/P_m$, and the magnetic field lines; Middle
panel: the ratio of azimuthal and poloidal magnetic field strengths,
$log_{10}B^\phi/B_p$, and the magnetic field lines; Right panel: the
magnetic field strength, $log_{10}(B)$, and the magnetic field lines.
}
\label{f2}
\end{figure*}

\section{Computer simulations}
\label{simulations}

The simulations were carried out with 2D axisymmetric GRMHD code
described in Komissarov\shortcite{K04b}.  Since this code can deal
only with time-independent spacetimes we are forced to start from the
point where the central BH has already been formed inside the
collapsing star. In the presented model the BH mass
$M\sub{bh}=3M_{\sun}$ and its angular momentum parameter $a=0.9$.  The
mass density and the radial velocity of the collapsing star are
described by the free-fall model of Bethe\shortcite{bethe}
corresponding to $t=1$s since the onset of collapse (see equations in
Komissarov \& Barkov, 2007).  The parameter $C$ is set to 9
corresponding to most massive stars. This gives us the free-fall mass
rate $ \dot{M} \simeq 0.5 M_{\sun} \mbox{s}^{-1}$. On top of this we
endowed the free-falling plasma with angular momentum and poloidal
magnetic field.  The angular momentum distribution describes a solid
body rotation up to the cylindrical radius $\varpi=6300\,\mbox{km}$.
Further out the angular momentum is constant,
$l=10^{17}\mbox{cm}^2\mbox{s}^{-1}$.  The magnetic field distribution
is that of a uniformly magnetized sphere in vacuum, the radius of this
sphere $r_1=4500$km and inside the sphere $B=3\times10^{10}$G. These
distributions are intended to describe the progenitor at the onset of
collapse rather then at the state developed one second later.  We
utilize the Kerr-Schild coordinates of spacetime.  The computational
grid is uniform in polar angle, $\theta$, where it has 180 cells.  In
the radial direction it is uniform in $\log r$, and has 450 cells.
The inner boundary is located just inside the event horizon and adopts
the free-flow boundary conditions. The outer boundary is located at
$r=25000km$ and at this boundary the flow is prescribed according to
the Bethe's model.

At the beginning of simulations the angular momentum of accreting gas
is less than that of the last stable orbit, $l\sub{lso}$. It falls
straight into the BH, and no disk is formed. However, the magnetic
flux threading the BH gradually increases and so is the magnetic
pressure. When the outer layers with $l>l\sub{lso}$ reach the BH the
centrifugal force slows down their infall and the accretion disk is
beginning to form around the BH. At the same time the accretion shock
separates from its surface.  The low angular momentum plasma of polar
regions k.png falling straight into the BH after passing the accretion
shock whereas the high angular momentum plasma fills the ``bubble''
above and below the disk (fig.\ref{f0}). Strong differential rotation
within this bubble leads to amplification of the azimuthal component
of magnetic field, the magnetic pressure grows and eventually
overwhelms the ram pressure of free-falling envelope -- the explosion
begins. The BH is a key player in the process pumping electromagnetic
energy into the bubble and the disk at the rate of $\simeq 2\times
10^{51}\mbox{erg}\,\mbox{s}^{-1}$.

Figures \ref{f1} and \ref{f2} show the solution at $t=0.45$s, near the
end of simulations. At this time, the solution exhibits two well
defined polar jets surrounded by the magnetic cocoons of high pressure
and low density.  The magnetic pressure of these cocoons, which have
been inflated by the jets, exceeds by more than six orders of
magnitude the magnetic pressure in the collapsing star.  These
over-pressured cocoons drive strong bow shock (blast wave) into the
star (right panel of fig.\ref{f1}). The mean propagation speed of the
shock in the polar direction $v\sub{s}\simeq 0.18c$.  Near the equator
the stellar plasma compressed by the shock continues streaming
downward with supersonic speed.  At the equator and well outside of
the accretion disk the stream coming from northern hemisphere collides
with the stream coming from the southern hemisphere and a pair of
oblique shocks develop at $r\simeq 200r_g$ (middle panel of
fig.\ref{f2}).  These shocks are not strong enough to cause
photo-dissociation of nuclei and the high post-shock pressure drives
the reflected flows away from the equatorial plane. Plasma from the
skin layers of the reflected streams actually enters the bubbles and
interacts with the jets (We expect this effect to weaken later when
the blast wave moves further away.)  The inner layers of the reflected
streams pass through another shock, at $r\simeq 50r_g$, and feed the
accretion disk.  The left panel of fig.\ref{f1} shows the solution in
the immediate vicinity of the BH. Its structure is reminiscent to that
found in the previous studies of thick disks around BHs -- main disk,
its dynamic corona, and magnetically-dominated
funnel~\cite{DH03,MG04,SST07}.  This funnel is the region there the
Pointing dominated jets are produced as well as the ``wind'' blowing
into the BH -- in this image one can clearly see the surface
separating these flows.

Figure \ref{f2} shows the magnetic properties of the central region.
Not only the funnel but also the disk corona are
magnetically-dominated.  The magnetic field strength reaches
few$\times10^{15}$G near the BH, it is weaker in the funnel compared
to the disk at the same spherical radius but not by much. Within the
disk and corona the azimuthal magnetic field, $B^\phi$ exceeds the
poloidal one, $B_p$, by two or three orders of magnitude. In contrast,
in the funnel $B^\phi/B_p \le 1$, reaching unity only near the funnel
walls.  In fact, the poloidal field in the funnel exceeds that in the
disk and corona by 1-2 orders of magnitude. This is in contrast to the
conclusion made by Ghosh \& Abramowicz\shortcite{GA97} and Livio et
al.\shortcite{LOP99}, namely that the poloidal field threading the BH
horizon should be of the same order as the poloidal field in the inner
parts of the disk. Their main argument, that both fields are produced
by the same azimuthal current flowing in the disk, misses the fact
that additional currents may flow in the magnetosphere and over the
disk/funnel surface and support the magnetic field inside the funnel
in the manner similar to solenoid.  In our case, the poloidal field
threading the BH is the original field of the progenitor that has been
accumulated during the initial phase of free infall.
 
\begin{figure*}
\includegraphics[width=57mm,angle=0]{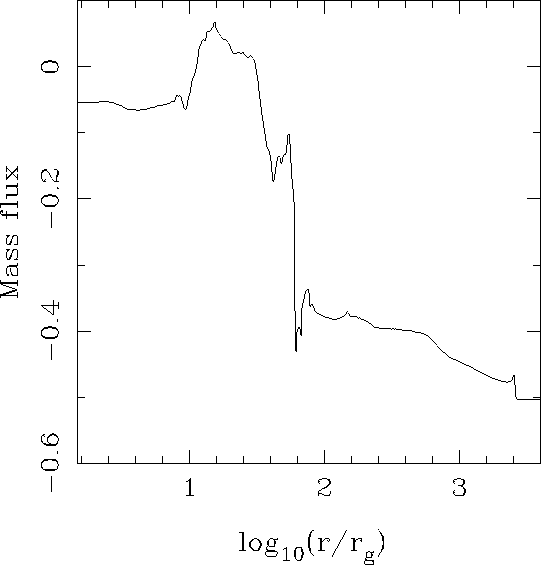}
\includegraphics[width=57mm,angle=0]{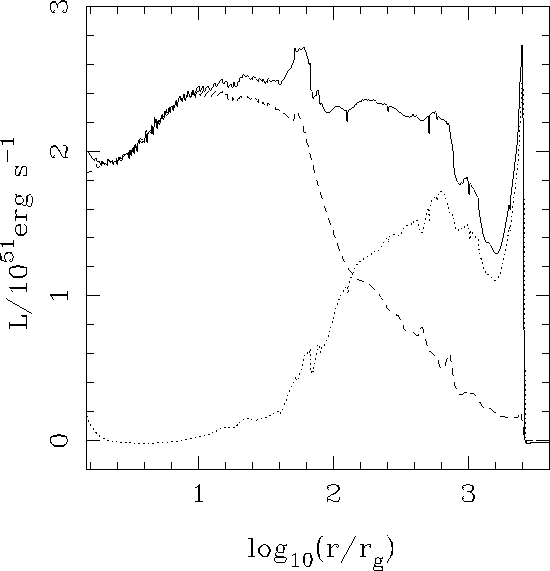}
\includegraphics[width=57mm,angle=0]{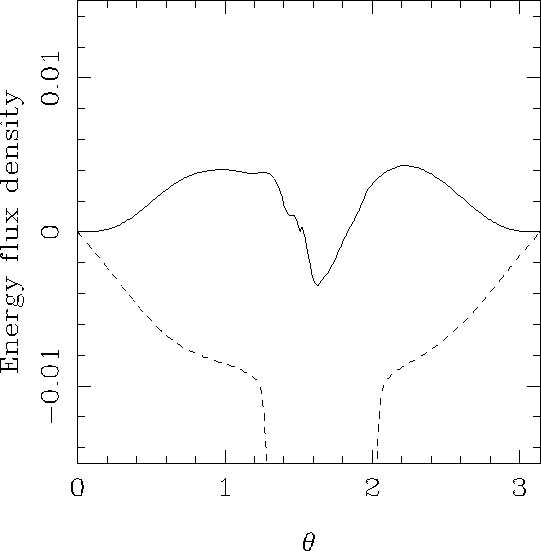}
\caption{Left panel: the integral baryonic mass flux in units
$M_{\sun}\mbox{s}^{-1}$ as a function of spherical radius; Middle
panel: the integral fluxes of total energy (solid line),
electromagnetic energy (dashed line), and hydrodynamic energy (dotted
line); Right panel: the energy flux densities at the event horizon for
electromagnetic energy (solid line) and hydrodynamic (matter)
energy. Time $t=0.45$s.}
\label{f3}
\end{figure*}

The left panel of fig.\ref{f3} shows the baryonic mass flux as a
function of spherical radius. One can see that it reduces from the
free-fall value $\dot{M}\simeq -0.5M_{\sun}\mbox{s}^{-1}$ down to
$\dot{M}\simeq -0.06M_{\sun}\mbox{s}^{-1}$ at the event
horizon. Between $r\simeq 60r_g$ and $r=2500r_g$ this reduction
reflects the effect of the bow shock driven into the star by the
jets. The sharp reduction at $r\simeq 60r_g$ corresponds to the
position of the accretion shock and marks the transition from
approximate free-fall to the centrifugally supported disk.

The middle panel of fig.\ref{f3} shows the integral energy fluxes of
the jets as functions of spherical radius. To be more precise the
integration is carried out over the the whole sphere but the
contribution from areas with the baryonic rest mass density
$\rho>10^8\mbox{g}\,\mbox{cm}^{-3}$ is excluded. We have verified that
the bulk contribution to the fluxes computed in this way comes from
the jets. The baryonic rest mass flux, $\rho u^r$ 
radial component of 4-velocity, is excluded from the total and the
matter energy fluxes, that is these fluxes are computed via

\begin{equation}
 \dot{E}=-2\pi\int_S (T^r_t + \rho u^r)\sqrt{\gamma} d\theta,
\end{equation}   
where $\gamma$ is the determinant of the metric tensor of space and
$\bT$ is either the total stress-energy-momentum tensor or its
hydrodynamic part.  The most important conclusion suggested by this
figure is that at least $80\%$ of the jet energy is provided directly
by the BH and at a very high rate, $\dot{E}\simeq
2\times10^{51}\mbox{erg}\,\mbox{s}^{-1}$. The remaining $20\%$ seem to
be provided by the inner part of the disk -- this explains the rise of
jet power between the event horizon and $r\simeq10r_g$.  Indeed,
careful examination of the solution shows that some magnetic field
lines enter the jet from the skin layers of the disk with
$\rho>10^8\mbox{g}\,\mbox{cm}^{-3}$. However, it remains to be shown
that this is not caused by the numerical diffusion of magnetic flux
from the funnel into the disk.  The right panel of fig.\ref{f3} shows
the distributions of Poynting flux and hydrodynamic energy flux
(including the rest mass-energy) across the horizon and allows us to
determine whether it is the Blandford-Znajek or the MHD-Penrose
mechanism \cite{PC90,Pun01,KSKM02} or both of them that provide the
energy supply to the jets.  Since the hydrodynamic flux is everywhere
negative the MHD-Penrose mechanism can be ruled out with
certainty. This is confirmed by the fact that the hydrodynamic
energy-at-infinity is positive everywhere inside the ergosphere. Thus
the jet is powered by the Blandford-Znajek mechanism. For a force-free
monopole magnetosphere the Blandford-Znajek power is given by

\begin{equation}
 \dot{E}\sub{BZ}=\frac{1}{6c}\left(\frac{\Omega_h\Psi}{4\pi}\right)^2,
\end{equation}
where $\Omega_h$ is the angular velocity of the BH and $\Psi$ is the
magnetic flux threading the BH. In the derivation we assumed that the
angular velocity of magnetosphere $\Omega =0.5\Omega_h$. This holds
well even for rapidly rotating BHs with monopole magnetospheres
\cite{K01} and corresponds to the mean value of $\Omega$ measured in
our simulations as well.  Using the measured value of $\Psi$ we derive
$\dot{E}\sub{BZ}\simeq 2.6\times10^{51}\mbox{erg}\,\mbox{s}^{-1}$
which agrees quite well with the value of $\dot{E}\sub{BZ}$ provided
by fig.\ref{f3}.  The total amount of free energy-at-infinity in the
bow shock and the bubble at time $t=0.45$s is $E\simeq 1.37\times
10^{51}\mbox{erg}$. Since the explosion develops only at $t=0.24$s the
mean jet power over the active period is $<\dot{E}>\simeq
6\times10^{51}\mbox{erg}\,\mbox{s}^{-1} $, indicating the higher jet
power at the early stages of the explosion.

The middle panel of fig.\ref{f3} also shows that initially the jets
are Poynting-dominated but gradually the electromagnetic energy is
converted into the energy of matter.  However, the accuracy of our
simulations is insufficient to capture the jet dynamics.  First of
all, we are forced to keep the flow magnetization below the limit at
which the code crashes -- this is done via artificial injection of
plasma in the danger cells.  This reduces the length scale for the
energy conversion via magnetic acceleration of plasma, as well as the
asymptotic Lorentz factor. The numerical mass diffusion into the jets
from the disk corona further exacerbates this problem.  Finally,
numerical resistivity causes dissipation of the jet electric
current. Due to the mass diffusion and numerical viscosity the jets
never become ultrarelativistic - their Lorentz factor rarely exceeds
$\Gamma=3$.  On the other hand, the total energy is conserved we do
not expect these numerical problems to have strong effect on the
dynamics of the bow shock and the bubble inflated by the jets.

\section{Conclusions}

Our results provide strong support to the idea that magnetic fields
can play a crucial role in driving powerful GRB jets and associated
stellar explosions not only in the magnetar model but also in the
collapsar model. The main energy source for the jets and explosions in
our simulations is the rotational energy of black hole and it is
released via the Blandford-Znajek mechanism. The measured rate of
energy release, $\dot{E} \ge 2\times10^{51}\mbox{erg}\,\mbox{s}^{-1}$,
can explain the energetics of even the shortest of long duration GRBs.
The fact that the rotational energy of black hole,
$E\sub{bh}\simeq\mbox{few}\times10^{53}\mbox{erg}$, exceeds the
typical explosion values derived from observations, $E\simeq
10^{52}\mbox{erg}$, suggests a self-regulating process in which the
black hole activity ceases when the blast wave terminates further mass
supply to the accretion disk.  The full details of the simulations
together with the results of parameter study will be presented
elsewhere.

\section*{Acknowledgments}
We thank G.S.Bisnovatyi-Kogan for helpful discussions of this problem.
This research was funded by PPARC under the rolling grant
``Theoretical Astrophysics in Leeds''.


\end{document}